\documentstyle[12pt,aasms4]{article}

\input epsf.sty
\input{psfig.tex}

\newcommand{\etal}{et~al.\ }
\newcommand{\eg}{e.g.\ }
\newcommand{\ie}{i.e.\ }
\newcommand{\Msun}{M$_{\odot}$}

\newcommand{\kms}{km~s$^{-1}$}
\newcommand{\Fefs}{$^{56}$Fe}
\newcommand{\Cofs}{$^{56}$Co}
\newcommand{\Nifs}{$^{56}$Ni}
\newcommand{\SiI}{Si~{\sc i}}
\newcommand{\SiII}{Si~{\sc ii}}
\newcommand{\SI}{S~{\sc i}}
\newcommand{\SII}{S~{\sc ii}}
\newcommand{\CaI}{Ca~{\sc i}}
\newcommand{\CaII}{Ca~{\sc ii}}
\newcommand{\FeI}{Fe~{\sc i}}
\newcommand{\FeII}{Fe~{\sc ii}}
\newcommand{\FeIII}{Fe~{\sc iii}}
\newcommand{\CoI}{Co~{\sc i}}

\newcommand{\CoIII}{Co~{\sc iii}}
\newcommand{\NiI}{Ni~{\sc i}}

\newcommand{\NiIII}{Ni~{\sc iii}}
\newcommand{\kopt}{\kappa_{\rm opt}}
\begin{document}

\title{Can differences in the nickel abundance in Chandrasekhar mass models 
explain the relation between brightness and decline rate of normal 
Type Ia Supernovae? }

\author{Paolo A. Mazzali$^{1,3}$, Ken'ichi Nomoto$^{2,3}$, 
Enrico Cappellaro$^4$, \\
Takayoshi Nakamura$^2$, Hideyuki Umeda$^2$, Koichi Iwamoto$^5$}
 
\altaffiltext{1}{Osservatorio Astronomico, Via Tiepolo, 11, Trieste, Italy}

\altaffiltext{2}{Department of Astronomy,
School of Science, University of Tokyo, Tokyo, Japan}

\altaffiltext{3}{Research Center for the Early Universe,
School of Science, University of Tokyo, Tokyo, Japan}

\altaffiltext{4}{Osservatorio Astronomico, vicolo dell'Osservatorio, 5, 
Padova, Italy }

\altaffiltext{5}{Department of Physics, College of Science and Technology, 
Nihon University, Tokyo, Japan}

\begin{abstract}
The use of Type Ia supernovae as distance indicators relies on the
determination of their brightness. This is not constant, but it can be
calibrated using an observed relation between the brightness and the properties
of the optical light curve (decline rate, width, shape), which indicates that
brighter SNe have broader, slower light curves. However, the physical basis for
this relation is not yet fully understood. Among possible causes are different 
masses of the progenitor white dwarfs or different opacities in
Chandrasekhar-mass explosions. We parametrise the Chandrasekhar-mass models
presented by Iwamoto \etal (1999), which synthesize different amounts of \Nifs,
and compute bolometric light curves and spectra at various epochs. Since
opacity in SNe~Ia is due mostly to spectral lines, it should depend on the 
mass of Fe-peak elements synthesized in the explosion, and on the temperature
in the ejecta. Bolometric light curves computed using these prescriptions for
the optical opacity reproduce the relation between brightness and decline rate.
Furthermore, when spectra are calculated, the change in colour between maximum
and two weeks later allows the observed relation between $M_B(Max)$ and $\Delta
m_{15}(B)$ to be reproduced quite nicely. Spectra computed at various epochs
compare well with corresponding spectra of spectroscopically normal SNe~Ia
selected to cover a similar range of $\Delta m_{15}(B)$ values.    
\end{abstract}

\section{Introduction} 

The use of Type Ia Supernovae (SNe~Ia) to probe cosmological parameters relies
on the possibility to calibrate the absolute magnitude of individual SNe from
the observed luminosity evolution (for a review see, \eg, Branch 1998).
Pskovskii (1977) first suggested that brighter SNe~Ia decline more slowly than
dimmer ones. This was later confirmed by Phillips (1993). Using modern data for
a small sample of well observed local SNe~Ia he showed that the relation exists
in the $B$, $V$ and $I$ bands, but it is steepest in $B$. The relation between
maximum $B$ brightness and the number of magnitudes the $B$ band declines in
the first 15 days after maximum, a quantity called $\Delta m_{15}(B)$, is
commonly referred to as the Phillips' relation. Hamuy \etal (1995, 1996b)
confirmed the relation using a bigger sample of objects, although their slopes
were flatter than Phillips' original values.

Riess et al. (1995) used a more sophisticated approach based on the analysis 
of the entire early light curve, which allows the SN maximum brightness to be
determined by comparison with a set of template light curves of objects whose
distance and reddening are assumed to be known. However, the absolute
magnitude - light curve shape relation is not yet calibrated exactly, as is 
shown by the differences among recent work (Riess et al. 1995, 1998).

Although the relation between SN~Ia brightness and decline rate has been used
extensively, its physical bases are not yet understood. Several groups have
published synthetic light curves obtained from explosion models which differ 
in many of their properties (progenitor WD mass, mode of the explosion). These
light curves span a wide range of maximum brightness and decline rates, and
although there are models that appear to reproduce some of the observations, 
a fully consistent picture is still lacking. In this paper we explore the
possibility that a series of Chandrasekhar-mass explosion models, which differ
essentially in the amount of \Nifs\ they synthesize, can explain the observed
range of light curves and spectra of at least {\em spectroscopically normal}
SNe~Ia.

\section{Basic Light Curve Physics} 

SNe Ia synthesize significant amounts of radioactive \Nifs, and their light
curves are powered by the deposition in the expanding SN ejecta of the
$\gamma$-rays and positrons produced by the decay chain \Nifs $\rightarrow$
\Cofs $\rightarrow$ \Fefs.

The shape of the light curve of a SN~Ia near maximum depends essentially on the
fact that optical photons emitted upon the deposition and thermalization of the
$\gamma$-rays and positrons do not immediately escape from the SN. These
photons must in fact first propagate through the optically thick SN ejecta,
where they interact with spectral lines and free electrons until they are
redshifted into a region of the spectrum where the opacity is low and they can
escape. Since the opacity is dominated by line processes, these regions
correspond to wavelengths where line opacity is low. This gives rise to the
very characteristic SN Ia spectrum, with broad absorption features and few
P-Cygni emissions, which correspond to these 'opacity windows' (\eg Pinto \&
Eastman 2000, Mazzali 2000).

Since it takes photons a finite time to emerge from the ejecta in what is
essentially a random walk process, very soon after the explosion the SN
luminosity is lower than the energy input into the ejecta. Maximum light occurs
when the instantaneous rates of deposition of hard radiation and emission of
optical light are roughly equal (Arnett 1982).  As time goes on, the delay
between energy deposition and emission of optical radiation becomes smaller and
smaller. At late times, in the so-called nebular phase, $\gamma$-ray deposition
becomes less efficient, and a significant contribution to the light curve is
made by the positrons, which are supposed to deposit {\em in situ} if a weak
magnetic field is present in the ejecta. Positrons carry only about 3.5\% of
the total decay energy. 

The properties of the peak of a SN~Ia light curve have been studied
analytically by Arnett (1982, 1996). The basic features are:
\begin{enumerate}
\item The brightness of the light curve at maximum is proportional to the mass 
of synthesized \Nifs.
\item The width of the light curve $\tau_{\rm LC}$ depends on the ejected mass, 
the kinetic energy of the explosion and the optical opacity as follows: 
\begin{equation} 
 \tau_{\rm LC} \propto \kopt^{1/2} M_{\rm ej}^{3/4} E_{\rm K}^{-1/4}.
\end{equation}
\end{enumerate}
Here $M_{\rm ej}$ is the ejected mass (\ie the WD mass, since SNe~Ia are not
supposed to leave a remnant behind), $E_{\rm K}$ is the kinetic energy of 
explosion and $\kopt$ is the grey opacity to optical photons.

\section{Observational facts and interpretation} 

That SNe~Ia do indeed synthesize different amounts of \Nifs\ is an established
observational fact. Spectroscopically `normal' SNe~Ia produce roughly 0.5
\Msun, but extreme cases range from 0.1 \Msun\ for SN~1991bg (Filippenko \etal
1992) to about 1 \Msun\ for SN~1991T (Spyromilio \etal 1992). These extreme
cases are spectroscopically peculiar SNe. However, even among `normal' SNe~Ia a
distribution of properties is observed (Nugent \etal 1995, Fisher \etal 1995),
which is quite likely the result of different synthesized \Nifs\ masses 
(Mazzali \etal 1998, Contardo \etal 2000). Can these differences in the mass of
\Nifs\ influence not only the peak brightness, but also the width of the light
curves?

One possibility is to assume that all SNe~Ia are self-similar events, but that
the progenitor masses, hence $M_{\rm ej}$, are different, so that all other
parameters (\Nifs\ mass, $E_{\rm K}$) scale accordingly. This could explain the
observations in a rather straightforward manner.  However, this interpretation
implies that sub-Chandrasekhar events, and possibly super-Chandrasekhar ones
(\eg SN~1991T, Fisher \etal 1999) are very common, since such events would be
necessary to explain not only spectroscopically peculiar events, but also the
faint end of the `normal' SN~Ia sequence (see \eg Cappellaro \etal 1997). This
is currently not a favourite scenario among the `explosive' community, since
sub-Chandrasekhar explosions that have the correct element distribution are
difficult to design (e.g., Hillebrandt \& Niemeyer 2000). 

Another possibility is a change of $E_{\rm K}$ only. However, $E_{\rm K}$ is
produced in almost equal amounts by burning a given mass to NSE (\ie mostly to
\Nifs), or by incomplete burning to Intermediate Mass Elements (IME) such as Si
and S. Therefore, one can expect that those SNe which synthesize more \Nifs\
are not likely to have a much higher $E_{\rm K}$ than SNe which produce less
\Nifs, unless the total mass burned (to NSE or to IME's) is also larger. Since
in most models the WD is almost completely burned to either \Nifs\ or to IME's,
($\sim 1.2$--1.3 \Msun\ for a Chandrasekhar mass white dwarf, Nomoto \etal
1984, 1994, Woosley \& Weaver 1994) there is not much room for a large $E_{\rm
K}$ variation if the mass of the progenitor WD is constant. Also, the width of
the light curve depends only weakly on $E_{\rm K}$.

So, if we want to restrict ourselves to Chandrasekhar-mass explosions, a
variation in the opacity $\kopt$ is the most promising direction to follow. 

Many models exist of Chandrasekhar-mass explosions. Models differing in the
details of burning and flame propagation can produce different amounts of
\Nifs\ (usually in the range 0.4 -- 0.8 \Msun) and display different light
curves (see, e.g., H\"{o}flich \etal 1995, 1998 and references therein). 
Different masses of \Nifs\ lead naturally to different SN brightnesses.

Among the possible physical reasons that could make a SN~Ia explosion 
synthesize different amounts of \Nifs\ are a different value of the
deflagration speed (caused by a different buoyancy force) or of the density at
which the burning wave makes a transition from a deflagration to a detonation
(DDT). Umeda \etal (1999) suggested that a different C/O ratio in the
progenitor WD may cause a variation in the deflagration speed or in the
deflagration-to-detonation transition density $\rho_{\rm DDT}$.

Iwamoto \etal (1999, see also Brachwitz et al. 2000) computed the explosion
hydrodynamics and the nucleosynthesis for three representative cases of 
$\rho_{\rm DDT}$. In their models, the explosion starts as a slow deflagration,
leading to the synthesis of a small amount ($\sim 0.1$ \Msun) of Fe-group
isotopes ($^{54}$Fe, $^{56}$Fe, $^{58}$Fe, $^{58}$Ni). The exact ratios of
these isotopes depend on the value of the electron fraction $Y_e$, which in
turn depends on the central density and composition of the WD and on the flame
speed. Meanwhile the WD expands, electron capture decreases and as the burning
proceeds further out radioactive \Nifs\ is mostly synthesized. Eventually, the
deflagration enters the region of incomplete Si burning and explosive O
burning, where the transition to a detonation occurs. In the detonated region,
both \Nifs\ and IME's are produced. The amount of \Nifs\ produced in the
detonated layer (after DDT) is sensitive to $\rho_{\rm DDT}$.  For a higher
value of $\rho_{\rm DDT}$, the detonated regions reach higher temperatures
because of smaller specific heats, and so more \Nifs\ is produced and the SN~Ia
is brighter.  The WS15 series of models: DD1, DD2 and DD3 of Iwamoto \etal
(1999) synthesize 0.56, 0.69 and 0.77 \Msun\ of \Nifs, respectively, for values
of $\rho_{\rm DDT}$ ranging from 1.7 to $3.0\; 10^7$g~cm$^{-3}$. The three
models WS15 have the same slow deflagration speed (1.5\% of the sound speed),
thus the amount of \Nifs\ synthesized in the deflagration zone is the same and
the difference in the masses of \Nifs\ synthesized is due entirely to the
different $\rho_{\rm DDT}$.

An interesting property of the models presented by Iwamoto \etal (1999) is 
that the total mass of Fe-peak elements (mostly \Nifs) plus IME's is roughly
constant ($\approx 1.28 - 1.30$ \Msun), so that all models also have about the
same $E_{\rm K}$ ($1.33 - 1.43 \; 10^{51}$erg).  Note that such a small range
of $E_{\rm K}$ would introduce a change of only less than 2\% in $\tau_{\rm
LC}$. It is therefore safe to ignore these differences. Therefore both 
$E_{\rm K}$ and $M_{\rm ej}$ are eliminated from Eq.(1) and only the dependence 
of $\tau_{\rm LC}$ on $\kopt$ remains. This feature makes these models
particularly appealing for an investigation of their photometric and
spectroscopic properties. The idea is to verify whether such a spread in the
properties of the models can explain the observed spread of properties of at
least {\em spectroscopically normal} SNe~Ia.

The question is how can a different \Nifs/IME ratio affect $\kopt$. There seem
to be two ways that this can happen. One is that more \Nifs\ leads to more
heating and therefore to a larger $\kopt$ (Khokhlov \etal 1993, H\"oflich \etal
1996). However, since most of the \Nifs\ is not mixed, an increase in the total
production of \Nifs\ does not immediately translate into a higher temperature.
The other possibility stems from the consideration that the opacity in the
ejecta of a SN~Ia is dominated by line opacity of low ionization species.
Pauldrach \etal (1995) showed that the $\tau_{\rm e} = 1$ surface in W7-type
ejecta at an epoch of 25 days falls at $v \sim 5000$ \kms, which is
significantly less than the photospheric velocity at that epoch.  In the
conditions that apply in a SN~Ia near maximum, the time a photon spends
scattering in lines as it redshifts its way out of the expanding envelope is
much shorter than the time spent travelling from one line to the next redder
line (Pinto \& Eastman 2000, Mazzali 2000). The crossing of each line in
frequency space can thus be treated as a single scattering event, so that the
opacity depends simply on the number of active lines, which in turn depends on
which elements dominate the composition. In particular, low excitation ions of
Fe-group elements (\FeI--\FeIII, \CoI--\CoIII, \NiI--\NiIII\ etc.) have many
more lines (about a factor of 10) than low excitation ions of IME's
(\SiI--\SiII, \SI--\SII, \CaI--\CaII\ etc.). It may therefore not be
unreasonable to expect that the average opacity is higher in regions where the
abundance of the Fe-group elements is higher. On average, one can therefore
expect that SNe with a higher \Nifs/IME production ratio also have a higher
opacity.

\section{Explosion models} 

Since our aim was to test one basic aspect of the light curve, namely the
dependence on the composition of the ejecta, we chose to simplify the input as
much as possible. In particular, since the models published by Iwamoto \etal
(1999) have similar density distributions (because they have similar $E_{\rm
K}$, $\sim 1.3 - 1.4 \; 10^{51}$erg), we adopted the W7 (Nomoto \etal 1984)
distribution as an average for all models.  This is a reasonable approximation
because $\rho(v)$ does not change greatly among different Chandrasekhar-mass
explosion models, and W7 is a good representation of the typical $\rho(v)$.
However, if the mass of \Nifs\ translates more or less directly into luminosity
at maximum, the spread in \Nifs\ mass of the Iwamoto \etal (1999) models
corresponds to a spread of maximum brightness of only about 0.35 mag. This is
less than the observed spread of `normal' SNe~Ia ($\sim 0.6$ mag). Therefore,
we constructed models which produce 0.4, 0.6 and 0.8 \Msun\ of \Nifs,
respectively, corresponding to $\rho_{\rm DDT} = 1.3 - 3 \: 10^7$g~cm$^{-3}$
(Iwamoto \etal 1999; Umeda \etal 1999).

In order to do this we followed the properties of the WS15 DD series of models
by Iwamoto \etal (1999). In these models, \Nifs\ is synthesized outside of an
enclosed mass of $\sim 0.1$ \Msun, and it is the dominant element in the ejecta
until the IME region is reached. Models with different masses of \Nifs\ are
therefore characterized by a different outer extent of the \Nifs-dominated
shell. Therefore we placed the interface between the \Nifs\ and the 
IME-dominated regions at the mass coordinate appropriate to reach the required
value  of the \Nifs\ mass.  The abundance distribution of our parametrised
models is shown in Fig.1. Using only two main contributions to the opacity, \ie
binning all elements into either Fe-group or IME, is also a step towards
simplifying the models, and it is justified because none of the lighter ions
have such a complicated level structure as any of the dominant Fe-group ions. 
It should be noticed that the distribution of \Nifs\ in velocity space in the
simple models we constructed compares favourably with the velocity of the
\Nifs\ sphere derived by Mazzali \etal (1998) when fitting the width of the
\FeII] and \FeIII] emission lines in normal SNe~Ia at late times.

\section{Calculations} 

In order to test the properties of these explosion models we adopt a 2-step
approach. First we compute the {\em bolometric} light curves using a Monte
Carlo code (Cappellaro \etal 1997). Our input models consist of a
density-velocity distribution, and three basic contributions to the composition
are singled out. In fact, for each of the shells into which the ejecta are
divided in our calculation we input the initial \Nifs\ abundance, $X_{^{56}{\rm
Ni}}$, according to which $\gamma$-rays are injected in the appropriate shell
following the decay law, and the total abundance of the Fe-group elements,
$X_{\rm Fe-gp}$, which includes \Nifs\ and its daughter nuclei but also stable
isotopes such as $^{54}$Fe, $^{56}$Fe, and $^{58}$Ni. These contribute to the
opacity but are not sources of $\gamma$-rays.  The remaining fraction of a
shell's mass is attributed to IME's.

As a first approximation, we initially adopted a formula for $\kopt$
which depends only on the relative abundance of Fe-group and IME. Since on 
average the number of active lines in an Fe-group is about 10 times larger 
than in an IME ion, we define the optical opacity as: 
\begin{equation}
 \kopt = 0.25 X_{\rm Fe-gp} + 0.025 (1 - X_{\rm Fe-gp} )
 						~~~[\rm cm^2 g^{-1}].
\end{equation} 
The opacity is assigned a value in each of the shells into which the ejecta 
have been divided. With this formula, $\kopt$ ranges from 0.25 cm$^2$ g$^{-1}$ 
in regions where Fe-group dominates to 0.025 cm$^2$ g$^{-1}$ in regions where 
Fe-group is absent, and has a value 0.1375 cm$^2$ g$^{-1}$ if the composition
is 50\% Fe-group and 50\% IME. This is a reasonable representation of the value
of $\kopt$ (see Khokhlov \etal 1993, Figure 24). However, the bolometric light
curves computed with this formulation of the opacity peak too soon ($\approx
15$ -- 16 days) and drop too rapidly: $\Delta m(15)(Bol)$ for these light
curves ranges from 0.75 mag for model Ni08 to 0.85 mag for model Ni04. 
Typical risetimes of local SNe~Ia are about 19--20 days (Riess \etal 1999), 
and values of $\Delta m(15)(Bol)$ for normal SNe~Ia range from 0.9 to 1.2 mag 
(Contardo \etal 2000). 

The early rise and the slow decline of these synthetic light curves suggests 
that the opacity is underestimated before maximum and overestimated after 
maximum. Khokhlov \etal (1993) pointed out that the temperature has also a
significant influence on the opacity, which is rapidly reduced as the
temperature drops below 10$^4$ K, which is the effective temperature of
observed normal SNe~Ia around maximum. We therefore added a time-dependent 
term to the opacity equation to mimic the decrease of temperature, and 
hence of opacity, with time. We adopted the formula: 
\begin{equation}
 \kopt = 
 	\left[ 0.25 X_{\rm Fe-gp} + 0.025 (1 - X_{\rm Fe-gp} )\right] 
 	\left( \frac{t_d}{17} \right)^{-\frac{3}{2}}~~~[\rm cm^2 g^{-1}],
\end{equation} 	
but we limited the time-dependent term on the right to a maximum value of 2. 
With this formulation, the value of $\kopt$ for a region where both Fe-group
and IME have an abundance of 0.5 is 0.275 cm$^2$ g$^{-1}$ at 10 days and
earlier, 0.137 cm$^2$ g$^{-1}$ at 17 days and only 0.06 cm$^2$ g$^{-1}$ at 30
days.  This increased pre-maximum value should reflect both the higher
temperature and density at those epochs and the greater relevance of the
electron scattering opacity.

The basic properties of the synthetic bolometric light curves computed with
this formulation of the opacity are summarized in Table 1.  As expected, models
which produce more \Nifs\ are brighter and decline more slowly. In Figure 2 
the synthetic light curves are compared to the $uvoir$ light curves of 3
spectroscopically normal SNe which have different values of $\Delta m_B(15)$:
SN~1990N ($\Delta m_B(15)=1.07$ mag); SN~1994D ($\Delta m_B(15)=1.31$ mag) and
SN~1992A ($\Delta m_B(15)=1.47$ mag). The observed $uvoir$ light curves have
been shifted along the y-axis to obtain a best overall match to the synthetic
light curves, which are not meant to be detailed fits of these SNe.  The
$uvoir$ light curve of SN~1990N was computed using the Cepheid distance modulus
$m-M=32.03$ mag (Saha \etal 1997) and shifted upwards by 0.1 mag. The light
curve of SN~1992A was taken from Suntzeff (1996), who used an SBF distance
modulus $m-M=30.65$ mag, and shifted upwards by 0.8 mag, resulting in a
distance modulus $m-M=31.45$ mag, which is consistent with the GCLF distance
$m-M=31.35$ mag published by Della Valle \etal (1998). Both of these SNe are
therefore compatible with the 'long' distance scale.  Finally, the light curve
of SN~1994D was computed using the recently published GCLF distance modulus
$m-M=30.40$ mag (Drenkhahn \& Richtler 1999). However, this light curve had to
be shifted upwards by as much as 0.75 mag to match the light curve of model
Ni06, so that the distance used for the plot in Figure 2 is $m-M=31.15$ mag.
Although this appears to be inconsistent with the GCLF distance, we note that
SN~1994D had an SBF distance modulus $m-M=30.86$ mag (Hamuy \etal 1996a).
Typically, GCLF distance are larger than SBF ones, so the distance we have used
for Figure 2 may not be inconsistent with the 'long' distance scale. Also,
$\Delta m_B(15)$ of model Ni06 matches very closely the same quantity for
SN~1994D (see Table 2).

Although the synthetic bolometric light curves match the $uvoir$ ones
reasonably well, the decline of the bolometric light curves in the 15 days
following maximum is smaller than the observed $B$-band decline. However, the
observed $B$-band decline is faster than the observed $V$-band decline, and the
observed SN~Ia colour changes from $B-V \sim 0$ at maximum to $B-V \sim 0.6$ at
two weeks after maximum, so the fast decline of the $B$ magnitude must be at
least partially the result of a change towards the red of the colour of the
spectrum. Note that all light curves reach maximum about 18 days after the
explosion, although the brightest model reaches maximum somewhat later.

We used our Monte Carlo spectrum synthesis code (Mazzali \& Lucy 1993, Lucy
1999, Mazzali 2000) to compute synthetic spectra for the three models. The code
requires as input a hydrodynamical model of the explosion and values for the
luminosity $L$ and the photospheric velocity $v_{\rm ph}$ at an epoch $t$.  We
can take both $L$ and $v_{\rm ph}$ from the light curve calculations. However,
such calculations are known to give a poor representation of the photospheric
velocity, in particular at times past maximum (Fig. 3). At those epochs codes
overestimate the velocities, most likely because the approximation of a gray
photosphere is not valid at all wavelengths (Khokhlov \etal 1993, H\"oflich
\etal 1995, Iwamoto \etal 2000, Mazzali, Iwamoto \& Nomoto 2000). Therefore, 
we can use $v_{\rm ph}$ values from the light curve calculations to compute
synthetic spectra at maximum, but we have to rely on velocity information from
observed spectra to compute spectra at 15 days after maximum. We also computed
spectra at one week before maximum, for completeness, using both $L$ and
$v_{\rm ph}$ from the light curve calculations. For each synthetic spectrum
calculation, we used a ratio of Fe-group v. IME obtained by mixing the
compositions of the layers above the photosphere in the particular explosion
model. The relative abundances of the IME's are based on the explosion model 
W7, and for the Fe-group the decay of \Nifs\ into \Cofs\ and \Fefs\ was taken
into account. Stable Fe-group elements are buried deep in the ejecta, and they
do not influence the spectra.

The integrated photometry from the synthetic spectra was used to compute
$\Delta m_B(15)$. We obtained $\Delta m_B(15)$ values ranging from 1.1 to 1.5
mag.  These values are comparable to those of observed spectroscopically normal
SNe~Ia. The numbers for the three models are given in Table 2.

Finally, we compare our synthetic spectra for the epochs $-7$ days, maximum 
and $+15$ days with the spectra of the same SNe used in Figure 2 for the light
curve comparison. We use spectra taken at epochs as close as possible to those
of the models. Figure 4 shows the pre-maximum spectra, Figure 5 the spectra at
maximum light and Figure 6 the spectra at 15 days after maximum. The observed
spectra have been corrected for redshift but not for reddening, which is
however small. The values of $\Delta m_B(15)$ for the various SNe are given in
the figures. No attempt was made to obtain detailed fits to the spectra. The
synthetic spectra appear to resemble the observed ones. It is remarkable that
spectra obtained from such different models are actually not very different.
This is because at a given epoch the photosphere is located at lower velocities
in models that produce less \Nifs, at least after about day 10 (see Fig. 3),
but since the luminosity is also lower in these models the effective 
temperatures in the three models are similar. This is consistent with the
results of Nugent \etal (1995). Also, the abundances in the layers that form
the spectra, especially at and before maximum, are not much affected by the
deep layers, and so the spectra are relatively insensitive to the large
differences in \Nifs\ mass.

\section{Conclusions} 

Using a set of light curves and spectra computed for three representative
models we have shown that the differences in the observed properties of SNe~Ia
all having Chandrasekhar-mass progenitors but producing different amounts of
\Nifs\ reproduce the observed dispersion of properties of at least {\em
spectroscopically normal} SNe~Ia. Our calculation procedure is not fully
self-consistent because the photospheric velocities obtained from the light
curve calculations are not always reliable, especially after maximum, and we
had to use estimates from observed spectra. Nevertheless, the results are
encouraging. Calculations with a self-consistent code (\eg Nugent \etal 1997)
could give more reliable results.

The range of properties displayed by the models we have studied is sufficient
to encompass the observed properties of {\em spectroscopically normal} SNe~Ia. 
However, {\em peculiar} SNe~Ia, such as SNe 1991T and 1991bg, seem to be well
outside this range, especially from a spectroscopic point of view. For SN~1991T
there is spectroscopic evidence that \Nifs\ was present also in the outer part
of the ejecta (Mazzali \etal 1995), which would seem to require a different
explosion model (\eg Yamaoka et al. 1992). SN~1991bg, on the other hand, was
so faint and its light evolution so fast that it may be difficult to explain it
within the framework of Chandrasekhar-mass explosions. The fact that several
SNe very similar to either SN~1991T or SN~1991bg have been observed but that
there are essentially no examples of objects that might fill the gap in
properties between these very extreme cases and the range of `normal' SNe~Ia
may suggest that these may actually be individual subtypes. More observations
of SNe~Ia are necessary to address this issue.

{\bf Acknowledgements.} It is a pleasure to thank Peter H\"oflich for useful
discussions concerning the behaviour of the opacity. 
We are grateful to David Branch, the referee, for his 
remarks on an earlier version of this paper. 
This work has been supported in part by the Grant-in-Aid for
Scientific Research (12640233, 12740122) and COE research (07CE2002) of the
Japanese Ministry of Education, Science, Culture, and Sports in Japan.


\newpage
\begin{deluxetable}{cccccc}
\scriptsize
\tablenum{1}
\tablecaption{Parameters of the synthetic light curves}
\tablehead{\colhead{model} &
\colhead{\Nifs\ mass} & 
\colhead{t(max)} & 
\colhead{M(max)(Bol)} &
\colhead{M(+15)(Bol)} & 
\colhead{$\Delta m_{15}$(Bol)}     \nl
\colhead{(\Msun)} &
\colhead{(d)} &
\colhead{(mag)} & 
\colhead{(mag)} & 
\colhead{(mag)} }
\startdata
 Ni04 & 0.4 & 18.0 & -18.91 & -17.82 & 1.09 \nl 
 Ni06 & 0.6 & 18.0 & -19.14 & -18.17 & 0.97 \nl 
 Ni08 & 0.8 & 18.5 & -19.32 & -18.37 & 0.95 \nl
\enddata
\end{deluxetable}

\begin{deluxetable}{cccccc} 
\scriptsize
\tablenum{2}
\tablecaption{Parameters of the synthetic spectra} 
\tablehead{\colhead{model} & 
\colhead{M(max)($B$)} &
\colhead{$B-V$(max)} & 
\colhead{M(+15)($B$)} &
\colhead{$B-V$(+15)} & 
\colhead{$\Delta m_{15}(B)$}     \nl
\colhead{~} &
\colhead{(mag)} &
\colhead{(mag)} & 
\colhead{(mag)} & 
\colhead{(mag)} & 
\colhead{(mag)} }
\startdata
 Ni04 & -18.85 & 0.03 & -17.39 & 0.85 & 1.46 \nl
 Ni06 & -19.04 & 0.03 & -17.76 & 0.77 & 1.28 \nl
 Ni08 & -19.21 & 0.07 & -18.13 & 0.56 & 1.08 \nl
\enddata
\end{deluxetable}


\begin{figure}[t]
\epsfxsize=13.5cm 
\hspace{3.5cm}
\epsfbox{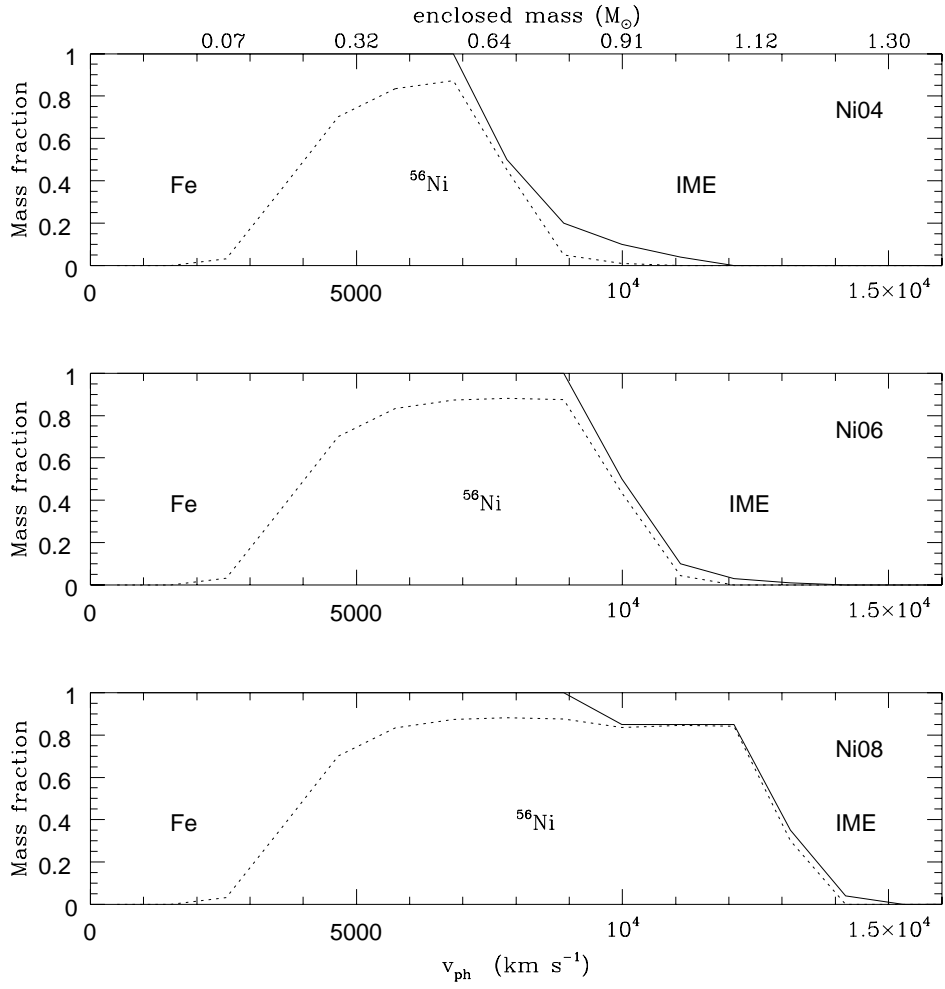}
\caption[h]{Composition of the three models.   }
\end{figure}

\begin{figure}[t]
\epsfxsize=13.5cm 
\hspace{3.5cm}
\epsfbox{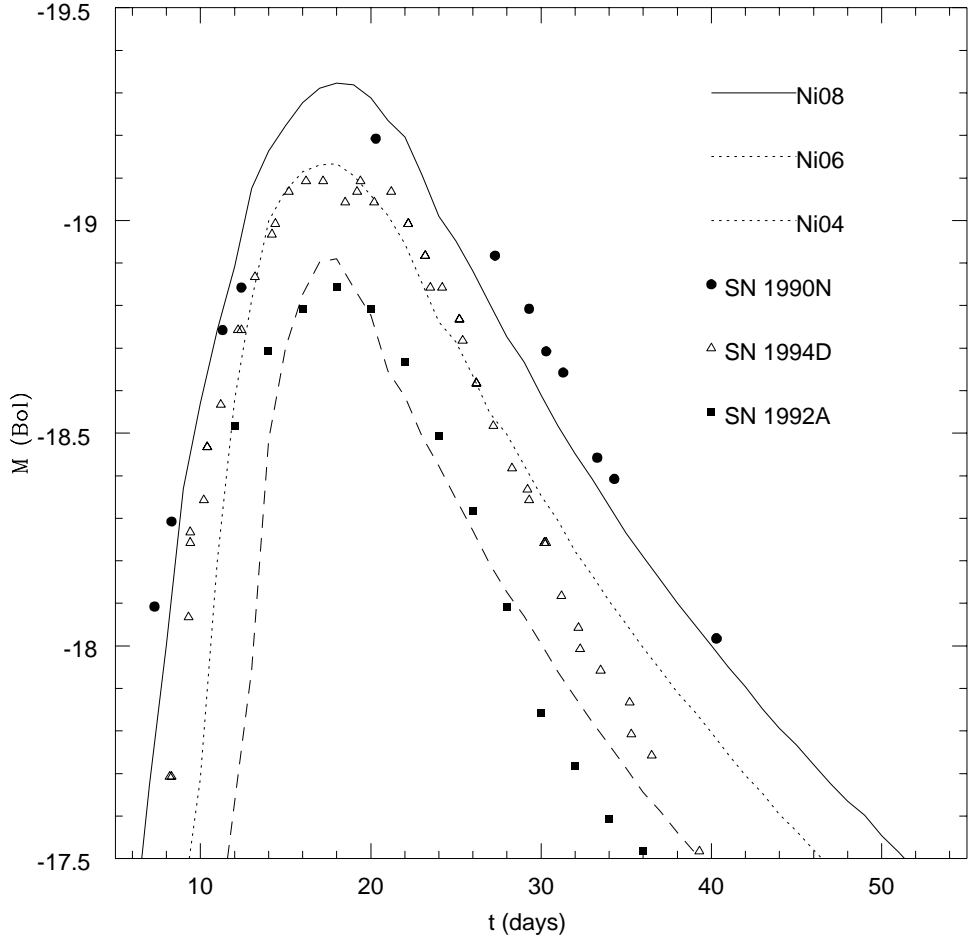}
\caption[h]{Synthetic bolometric light curves compared to $uvoir$ light curves
of spectroscopically normal SNe~Ia.   }
\end{figure}

\begin{figure}[t]
\epsfxsize=13.5cm 
\hspace{3.5cm}
\epsfbox{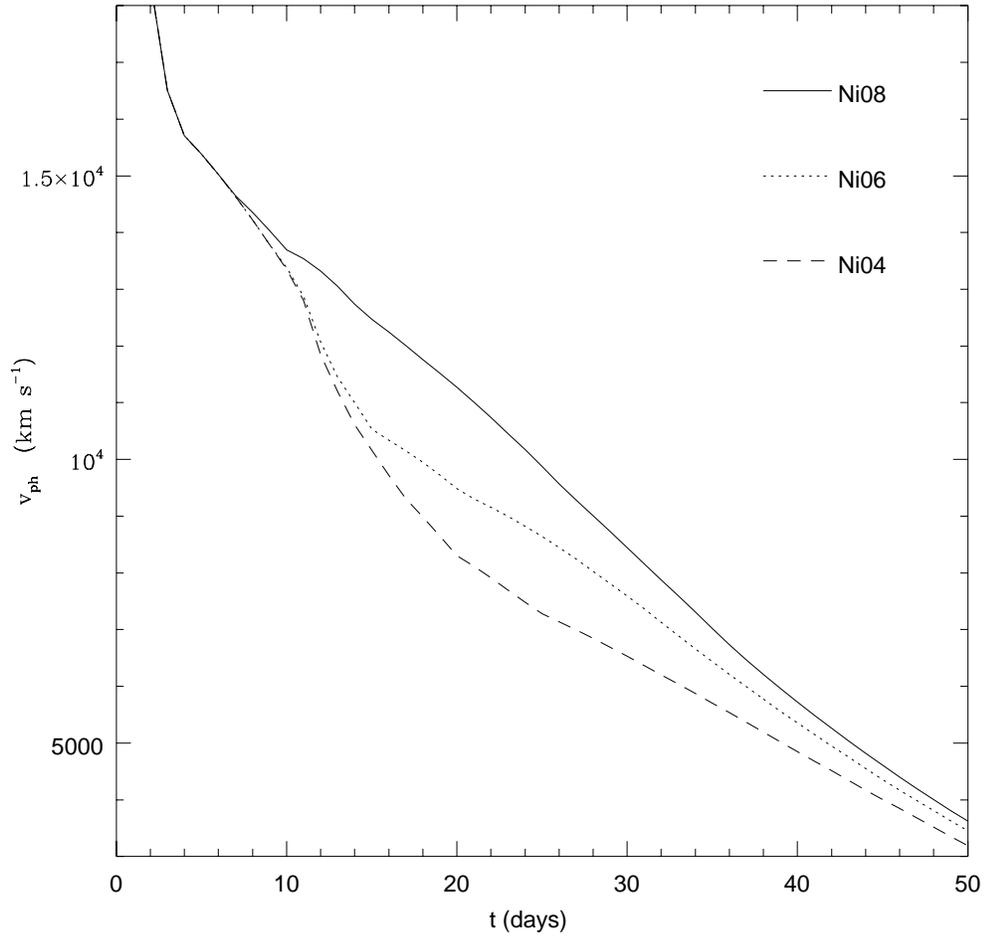}
\caption[h]{Photospheric velocities from the light curve calculations.   }
\end{figure}

\begin{figure}[t]
\epsfxsize=13.5cm 
\hspace{3.5cm}
\epsfbox{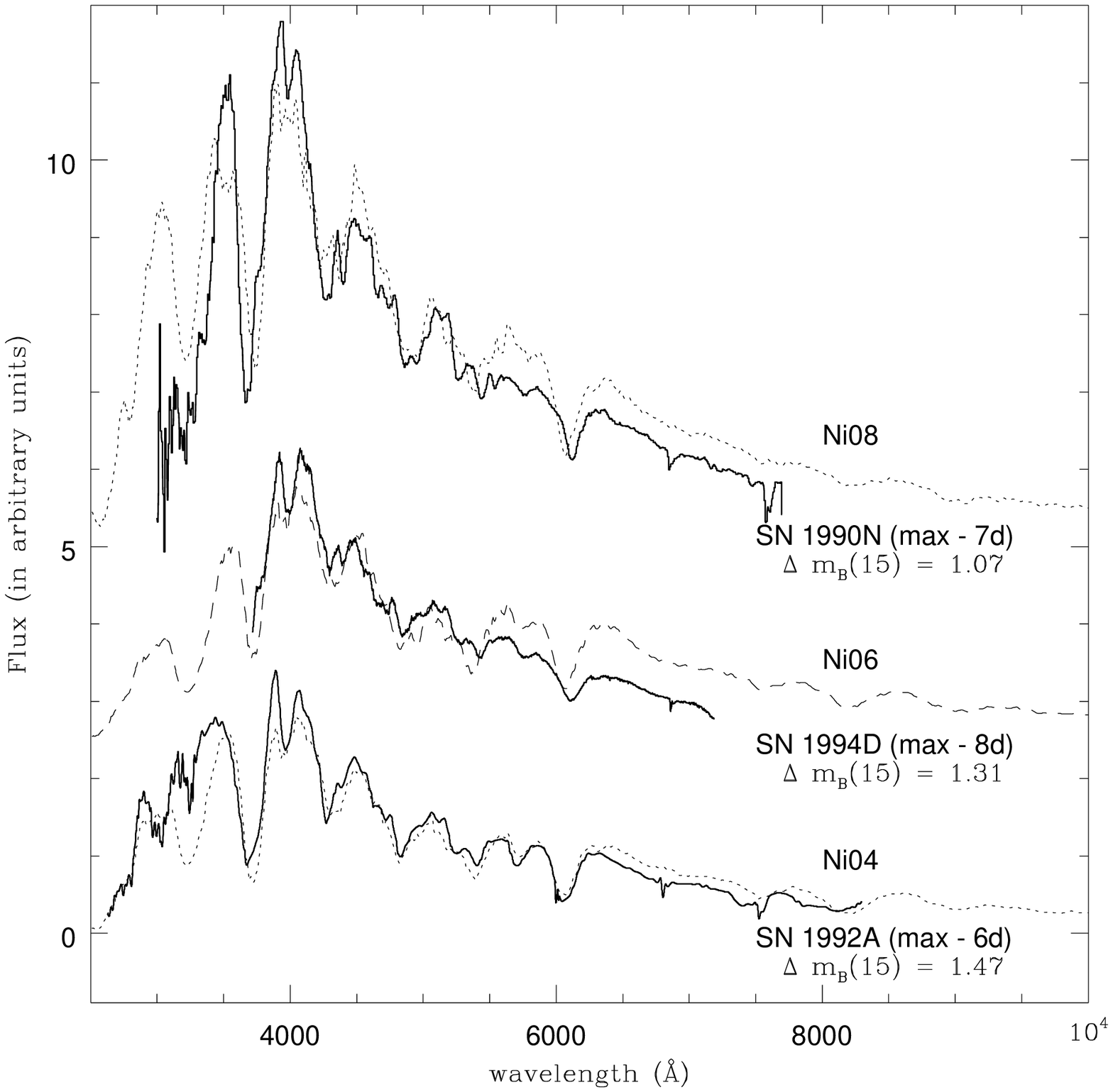}
\caption[h]{Synthetic and observed spectra at $t = Max - 7$ days.   }
\end{figure}

\begin{figure}[t]
\epsfxsize=13.5cm 
\hspace{3.5cm}
\epsfbox{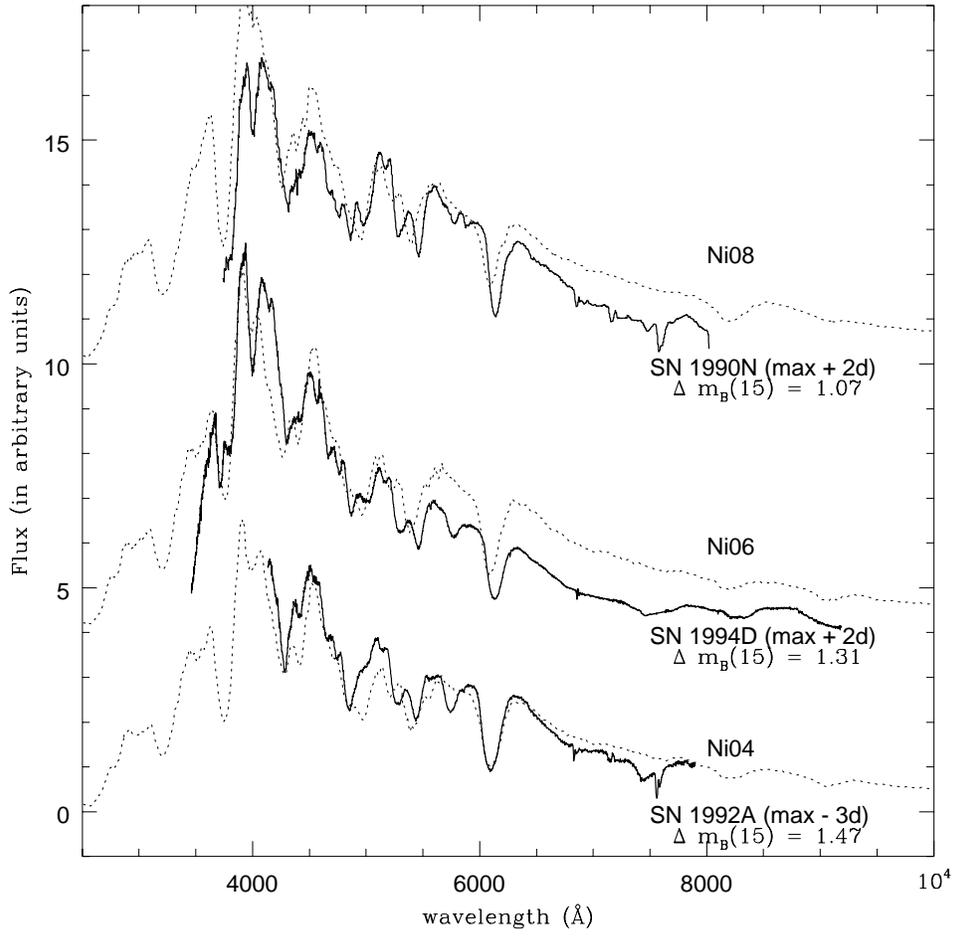}
\caption[h]{Synthetic and observed spectra at maximum.   }
\end{figure}

\begin{figure}[t]
\epsfxsize=13.5cm 
\hspace{3.5cm}
\epsfbox{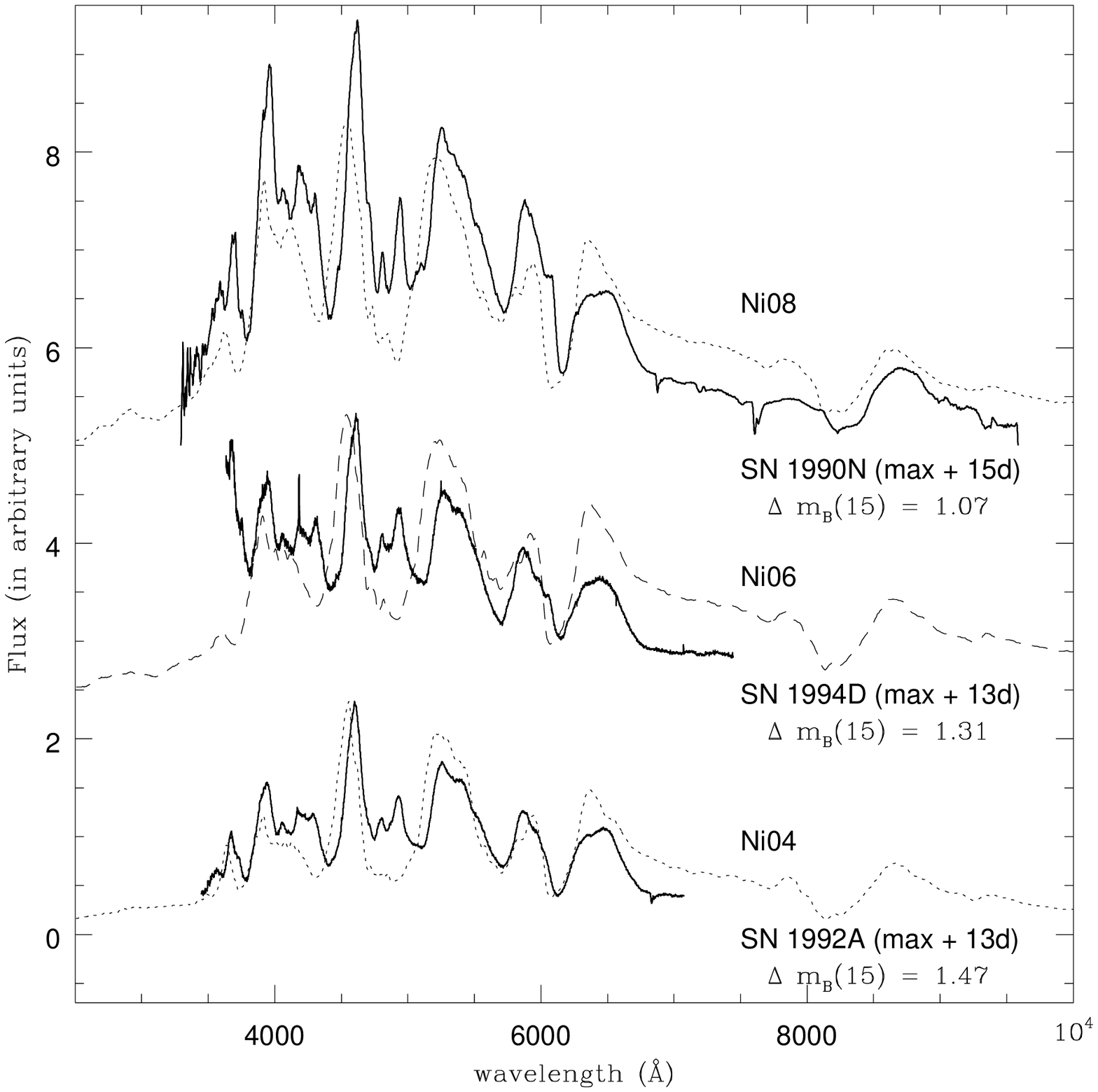}
\caption[h]{Synthetic and observed spectra at $t = Max +15$ days.   }
\end{figure}


\noindent
\begin{thebibliography}{}

\bibitem[Arn82]{arn82} Arnett, D., 1982, ApJ 253, 785 

\bibitem[Arn96]{arn96} Arnett, D., 1996,  Supernovae and Nucleosynthesis,
Princeton Univ. Press, Princeton

\bibitem[Bra00]{Bra00} Brachwitz, F., Dean, D.J., Hix, W.R., Iwamoto, K., 
Langanke, K., Martinez-Pinedo, G., Nomoto, K., Strayer, M.R., Thielemann, F.-K.,
Umeda, H. 2000, ApJ 536, 934
 
\bibitem[Bra98]{Bra98} Branch, D., 1998, ARAA 36, 17 

\bibitem[Cap97]{cap97} Cappellaro E., Mazzali P.A., Benetti S., et al. 1997, 
A\&A 328, 203

\bibitem[Cont00]{cont00} Contardo, G., Leibundgut, B., Vacca, W.D., 2000, A\&A
359, 876 

\bibitem[DV98]{dv98} Della Valle, M., Kissler-Patig, M., Danziger, I.J., 
Storm, J., 1998, MNRAS 299, 267

\bibitem[DR99]{dr99} Drenkhahn, G., Richtler, T., 1999 A\&A 349, 877

\bibitem[Fil92]{fil92} Filippenko,~A.V., Richmond,~M.W., Branch,~D., et al., 
1992, AJ 104, 1543 

\bibitem[Fis95]{fis95} Fisher,~A., Branch,~D., H\"{o}flich, P., Khokhlov, A.M., 
1995, ApJ 447, L73 

\bibitem[Fis99]{fis99} Fisher,~A., Branch,~D., Hatano, K., Baron, E., 1999, 
MNRAS 304, 67 

\bibitem[Ham95]{ham95} Hamuy, M., Phillips, M.M., Maza, J., Suntzeff, N.B.,
Schommerr~R.A., Aviles,~R., 1995, AJ 109, 1 

\bibitem[Ham96a]{ham96a} Hamuy, M., Phillips, M.M., Schommerr~R.A., 
Suntzeff, N.B., Maza, J., Aviles,~R., 1996a, AJ 112, 2391 

\bibitem[Ham96b]{ham96b} Hamuy, M., Phillips, M.M., Suntzeff, N.B., 
Schommerr~R.A., Maza, J., Aviles,~R., 1996b, AJ 112, 2398 

\bibitem[HN00]{hn00} Hillebrandt, W., Niemeyer, J., 2000, ARAA, in press 

\bibitem[Hof95]{hof95} H\"{o}flich, P., Khokhlov, A.M., Wheeler, J.C., 1995,
ApJ 444, 831 

\bibitem[Hof96]{hof96} H\"{o}flich, P., Khokhlov, A.M., Wheeler, J.C.,
Phillips,~M.M., Suntzeff,~N.B., Hamuy,~M., 1996, ApJ 472, L81 

\bibitem[Hof98]{hof98} H\"{o}flich, P., Wheeler, J.C., Thielemann, F.-K., 
1998, ApJ 495, 617 

\bibitem[Iwa99]{iwa99} Iwamoto, K., Brachwitz, F., Nomoto, K., et al., 1999, 
ApJS 125, 439 

\bibitem[Iwa00]{iwa00} Iwamoto, K., Nakamura, T., Nomoto, K., et al., 2000, 
ApJ 534, 660 

\bibitem[KMH93]{kmh93} Khokhlov, A., M\"uller, E., H\"oflich, P. 1993, 
A\&A 270, 223

\bibitem[Lucy]{Lucy} Lucy,~L.B. 1999, A\&A, 345, 211 

\bibitem[M2000]{M2000} Mazzali,~P.A., 2000, A\&A, in press

\bibitem[M98]{m98} Mazzali,~P.A., Cappellaro,~E., Danziger,~I.J., Turatto, M,
Benetti,~S. 1998, ApJ 499, L49 

\bibitem[ML95]{ML95} Mazzali,~P.A., Danziger,~I.J., Turatto,~M., 1995, A\&A
297, 509 

\bibitem[M00]{M00} Mazzali,~P.A., Iwamoto, K., Nomoto, K., 2000, ApJ, in press  

\bibitem[ML93]{ML93} Mazzali,~P.A. \& Lucy,~L.B. 1993, A\&A 279, 447

\bibitem[Nom84]{nom84} Nomoto,~K., Thielemann,~F.-K., Yokoi,~K., 1984, ApJ 286, 
644

\bibitem[Nom94]{nom94} Nomoto,~K., Yamaoka, H., Shigeyama, T., Kumagai, S,
Tsujimoto, T., 1994, in Supernovae, eds. S.A.Bludman et al.,  Amsterdam, 
Elsevier Science, 199 

\bibitem[Nug95]{nug95} Nugent,~P., Phillips,~M., Baron,~E., Branch,~D., 
Hautschild,~P.H. 1995, ApJ 455, L147

\bibitem[Nug97]{nug97} Nugent,~P., Baron,~E., Branch,~D., Fisher,~A., 
Hautschild,~P.H. 1997, ApJ 485, 812

\bibitem[Pau96]{pau96} Pauldrach~A.W.A., Duschinger~M., Mazzali~P.A., et al. 
1996, A\&A 312, 525

\bibitem[Phil]{phil} Phillips, M.M., 1993, ApJ 413, L105 

\bibitem[PE]{pe} Pinto, P.A., \& Eastman, R.G., 2000, ApJ 530, 757 

\bibitem[Psk77]{psk77} Pskovskii, Y.P., 1977, Sov. Astron. 21, 675 

\bibitem[Rie95]{rie95} Riess, A.G., Press, W.H., Kirshner, R.P., 1995, ApJ 438,
L17 

\bibitem[Rie98]{rie98} Riess, A.G., et al, 1998, AJ 116, 1009 

\bibitem[Rie99]{rie99} Riess, A.G., et al, 1999, AJ 118, 2675

\bibitem[Sah92]{sah92} Saha, A., Sandage, A., Labhardt, L., Tammann, G.A.,
Macchetto, F.D., Panagia, N., 1997, ApJ 486, 1 

\bibitem[Spy92]{spy92} Spyromilio,~J., Meikle,~W.S.P., Allen,~D.A., 
Graham,~J.R., 1992, MNRAS 258, 53P

\bibitem[Sun96]{sun96} Suntzeff,~N., 1996, in Supernovae and Supernova
Remnants, eds. R.McCray, Z. Wang, Cambridge, Cambridge Univ. Press, 41 

\bibitem[Ume99]{ume99} Umeda,~H., Nomoto,~K., Kobayashi,~C., Hachisu,~I.,
Kato,~M., 1999, ApJ 522, L43 

\bibitem[WW]{WW} Woosley,~S.E., \& Weaver,~T.E., 1994, in Supernovae, eds.
S.A.Bludman et al., Elsevier Science, Amsterdam, 63 

\bibitem[Ya92]{ya92} Yamaoka, H., Nomoto,~K., Shigeyama,~T., Thielemann,~F.K.,
1992, ApJ 393, L55

\end{thebibliography}
\end{document}